\documentclass[11pt]{article}

\usepackage[margin=1in]{geometry}
\usepackage{amsmath,amssymb,amsfonts}
\usepackage{graphicx}
\usepackage{booktabs}
\usepackage{natbib}
\usepackage{hyperref}
\usepackage{subcaption}
\usepackage[ruled,noend]{algorithm2e}

\title{Effective, Efficient, and General Information Abstraction for Imperfect-Information Extensive-Form Games}
\author{Boning Li\thanks{IIIS, Tsinghua University. Email: li-bn22@mails.tsinghua.edu.cn} 
\and Longbo Huang\thanks{IIIS, Tsinghua University. Email: longbohuang@tsinghua.edu.cn. Corresponding author.}}

\date{\today}

\begin{document}
\maketitle

\begin{abstract}
Information abstraction reduces the computational cost of solving imperfect-information games by clustering information sets into a smaller number of \emph{buckets}. Existing methods either rely on domain-specific features such as rank or equity, which are inapplicable to games with non-standard payoff structures, or require expensive offline neural-network training on billions of samples. We propose \textbf{Warm-up Expected Value-based Abstraction (WEVA)}, a simple yet effective alternative: run a small number of Counterfactual Regret Minimization (CFR) iterations on the full game as a \emph{warm-up} phase, extract per-hand expected value features at every decision node, form a depth-weighted multi-node feature vector, and apply $k$-means++ clustering to obtain the abstraction mapping. WEVA requires no domain knowledge, no pre-training, and incurs only a small overhead on top of the abstract-game solve. Experiments on three structurally diverse games, with different bucket numbers and CFR variants, show that WEVA consistently outperforms equity-based and rank-based abstractions, reducing exploitability by up to over $80\%$. Surprisingly, as few as $W{=}10$ warm-up iterations already produce abstractions that outperform existing information abstraction methods in most settings. These results establish WEVA as an \emph{effective, efficient, and general} approach to information abstraction in imperfect-information extensive-form games.
\end{abstract}

\section{Introduction}
\label{sec:intro}

Imperfect-Information Extensive-Form Games (IIEFGs) model sequential multi-agent decision-making under partial observability and provide a unifying framework for a wide range of domains, including poker \citep{bowling2015heads}, Mahjong \citep{liu2023opponent} and dark chess \citep{zhang2026general}.  Solving these games requires reasoning over information sets, i.e., collections of game states indistinguishable to a player, whose number grows exponentially with game size. Even with efficient algorithms like Counterfactual Regret Minimization (CFR) \citep{zinkevich2007regret} and its variants \citep{lanctot2009monte,brown2019solving,farina2021faster}, large games remain intractable without simplification. \emph{Information abstraction} addresses this by grouping similar information sets into \emph{buckets}, reducing the effective game size \citep{gilpin2006competitive}. This serves two purposes \citep{sandholm2010state,sandholm2015abstraction}: (1) it reduces computational cost by shrinking the strategy space, and (2) it provides an automatic classification of information sets that captures strategically relevant similarities.

Existing abstraction methods face a trade-off between generality and computational cost. Domain-specific features, such as \emph{rank} (relative hand strength) and \emph{equity} (expected win probability against a uniform opponent range) in poker \citep{shi2000abstraction,billings2003approximating,gilpin2006competitive}, are cheap to compute but tied to specific game structures: rank assumes a fixed hand ordering, and equity requires a well-defined showdown payoff matrix. When applied to non-standard games or games with non-monotonic payoff structures, these features degrade or fail entirely. Recent approaches  \citep{brown2019deep,li2025evpa} avoid domain dependency by training neural networks to estimate information set features, but require training on billions of samples, incurring substantial offline cost. We ask: 

\begin{center}
\framebox{%
  \parbox{0.85\textwidth}{%
    \centering
    \textit{Can we obtain EV-based abstraction features 
    cheaply, without neural networks or domain knowledge?}
  }%
}
\end{center}

\begin{figure}[!ht]
\centering
\includegraphics[width=\columnwidth]{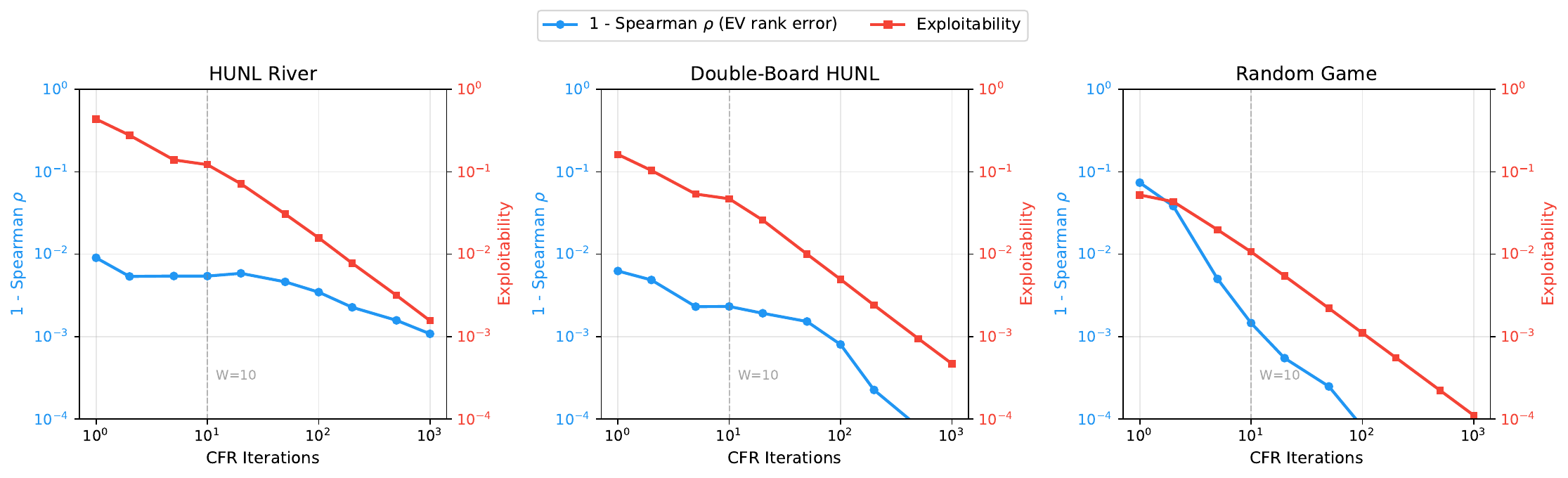}
\caption{Convergence of EV rank correlation (Spearman $\rho$) and exploitability during CFR iterations across different games.}
\label{fig:ev_convergence}
\end{figure}

A key observation motivates our approach. During CFR iteration, the \emph{ordinal ranking} of per-hand expected values stabilizes far earlier than the strategy itself converges (Figure~\ref{fig:ev_convergence}). We measure this using the Spearman rank correlation \citep{zar2005spearman} coefficient $\rho$, a standard measure of the monotonic association between two rankings: $\rho = 1$ indicates perfectly identical orderings, and $\rho = 0$ indicates no monotonic relationship. Comparing intermediate EVs against near-oracle EVs obtained by running $50{,}000$ CFR iterations, we find that $\rho$ exceeds 0.996 after only $W{=}10$ iterations across all three benchmark games, while exploitability remains orders of magnitude from convergence. In other words, CFR quickly learns \emph{which hands are strong and which are weak} long before it finds the equilibrium strategy. Since the relative ordering of EVs, which is the primary signal for grouping similar information sets, stabilizes early, the warm-up EV features can provide a reliable basis for similarity measurement.

Building on this observation, we propose \textbf{Warm-up Expected Value-base Abstraction (WEVA)}, a simple framework that extracts EV features directly from a short CFR warm-up phase. The method runs $W$ iterations of CFR on the full (unabstracted) game, extracts per-hand EV values at each decision node, constructs depth-weighted feature vectors, and applies $k$-means++ clustering \citep{DBLP:conf/soda/ArthurV07}. We compare two variants: \textbf{ev-root}, which uses only the root-node EV as a one-dimensional feature, and \textbf{ev-nd} (\emph{node-depth}), which constructs a multi-dimensional feature vector with exponentially decaying depth weights \citep{li2025evpa}. WEVA does not require domain knowledge, does not rely on neural network training, and incurs minimal abstraction overhead.

We evaluate WEVA on three structurally diverse river endgame benchmarks: Heads-up No-limit Hold'em (HUNL) Endgame, Double-Board HUNL (two independent boards with split pots), and a Random Game (per-node independent random payoffs). Across all three games, two state-of-the-art CFR variants (PCFR+ \citep{farina2021faster} and DCFR \citep{brown2019solving}), and multiple bucket counts ($K \in \{20, 50, 200\}$), WEVA consistently outperforms equity- and rank-based baselines, reducing exploitability by up to over $80\%$. Surprisingly, as few as $W{=}10$ warm-up iterations already suffice to beat all existing abstraction methods in most settings.

Our contributions are:
\begin{itemize}
    \item We propose \textbf{WEVA}, a lightweight EV-based information-abstraction framework that does not require domain knowledge or pre training. Our method is open-source and easily reproducible\footnote{Code is available at: https://github.com/lbn187/WEVA}, lowering the barrier for future research in information abstraction.

    \item Through experiments on three structurally diverse games with two CFR variants and three bucket settings, we show that WEVA consistently outperforms equity-based and rank-based baselines, reducing exploitability by up to over $80\%$.

    \item We show that even $W{=}10$ warm-up CFR iterations already produce abstractions that outperform existing methods in most settings, making WEVA an effective, efficient, and general information abstraction method in IIEFGs.

\end{itemize}

\section{Related Work}
\label{sec:related}


\paragraph{Information Abstraction.} Information abstraction groups similar information sets into buckets in a public state, which was initially performed manually in the poker domain via rank or equity \citep{shi2000abstraction,billings2003approximating}. Based on these domain features, automated abstraction was pioneered by \citet{gilpin2006competitive} for poker. ‌Subsequently, by utilizing the staged dealing property of poker, the next stage's bucket situation can be used as a better feature for clustering in flop and turn \citep{gilpin2007better,gilpin2007potential,gilpin2008heads}. In recent years, there have been some neural network-based abstraction methods \citep{brown2019deep,li2025evpa,kubicek2026lookahead,fu2026no}, which do not require domain knowledge but require a large amount of offline data and computational overhead for training. The strength of information abstraction is positively correlated with the scale of abstraction \citep{kroer2014extensive}, but it is not completely monotonous \citep{waugh2009abstraction}. Finding a universally applicable and efficient abstraction for various games remains difficult \citep{gilpin2008expectation,sandholm2015abstraction}, which is exactly the problem that our work aims to solve.

\paragraph{Other Abstractions.} Isomorphic abstraction is lossless  \citep{gilpin2006finding,gilpin2007lossless}, but its compression factor is limited and depends on game properties. The Imperfect-recall abstraction forces the agent to forget some of the public state information \citep{waugh2009practical,lanctot2012no}, thereby merging information sets from different public states and further reducing the size of the game tree. Imperfect-recall abstraction can be combined with information abstraction to generate high-quality abstractions for large-scale games such as HUNL \citep{johanson2013evaluating,ganzfried2014potential,brown2015hierarchical}. Action abstraction reduces the branching factor by selecting a subset of available actions at each decision point \citep{hawkin2011automated, hawkin2012using,brown2014regret}, which has been widely used in practical game-solving systems \citep{brown2018superhuman,li2024rl}.

\paragraph{CFR algorithms.}
CFR \citep{zinkevich2007regret} and its variants form the backbone of game-solving algorithms with theoretical guarantees.  
Discounted CFR (DCFR) \citep{brown2019solving} applies time-dependent discounting to past regrets and strategies. Predictive CFR+ (PCFR+) \citep{farina2021faster} further improves convergence by incorporating predictive regret matching. CFR can be combined with techniques such as sampling \citep{lanctot2009monte,johanson2011accelerating}, abstraction \citep{brown2015simultaneous}, pruning \citep{brown2015regret,brown2017reduced}, regularization \citep{perolat2021poincare}, 
warm-start \citep{brown2016strategy}, decomposition \citep{burch2014solving}, re-solving \citep{brown2017safe} and search \citep{brown2018depth,schmid2023student} to achieve efficient equilibrium findings.

\section{Notation}
\label{sec:notation}

An Imperfect-Information Extensive-Form Game (IIEFG) \citep{kuhn1953extensive} is defined by the tuple
$G = (\mathcal{N}, \mathcal{H}, \mathcal{A}, \mathcal{Z}, \mathcal{P}, u, \sigma_c, \mathcal{I})$.
$\mathcal{N} = \{1, 2\}$ denotes the two players (we focus on two-player zero-sum games).
$\mathcal{H}$ is the set of all histories (sequences of actions from the initial history $\emptyset$).
For any history $h \in \mathcal{H}$, $\mathcal{A}(h)$ is the set of available actions; executing $a \in \mathcal{A}(h)$ transitions to a new history $ha$.
$\mathcal{Z} \subseteq \mathcal{H}$ is the set of terminal histories, with $u_p(z)$ the payoff for player $p$ at $z \in \mathcal{Z}$.
The player function $\mathcal{P}: \mathcal{H} \setminus \mathcal{Z} \to \mathcal{N} \cup \{c\}$ specifies the acting player, where $c$ is the chance player following a fixed strategy $\sigma_c$.

For each player $p$, the information partition $\mathcal{I}_p$ groups indistinguishable histories: for any $I \in \mathcal{I}_p$ and $h, h' \in I$, player $p$ cannot distinguish $h$ from $h'$.
We write $I_p$ to emphasize that an information set belongs to player $p$; when the player is clear from context, we use $I$ and $I_p$ interchangeably.
We denote by $\mathcal{I}_p^0 \subseteq \mathcal{I}_p$ the set of \emph{root information sets}, i.e., the information sets at the game's initial decision point for player $p$. For a root information set $I_p^0 \in \mathcal{I}_p^0$ and a decision node $n$, we write $I_p^0{\downarrow}n$ for the \emph{downstream information set} at $n$: the unique information set sharing the same private information as $I_p^0$, combined with the public action history leading to $n$.
A behavioral strategy $\sigma_p$ assigns a probability distribution over $\mathcal{A}(I)$ for each $I \in \mathcal{I}_p$; the probability of action $a$ at $I$ is denoted $\sigma(I, a)$.
A strategy profile $\sigma = (\sigma_1, \sigma_2)$ determines the reach probability $\pi^{\sigma}(h)$ of any history $h$.
The counterfactual value for player $p$ at information set $I$ is
\[
v_p^{\sigma}(I) = \sum_{h \in I} \pi_{-p}^{\sigma}(h) \sum_{\substack{z \in \mathcal{Z},\, h \sqsubseteq z}} \pi^{\sigma}(z \mid h)\, u_p(z).
\]
And the expected value (EV) for player $p$ at information set $I_p$ under $\sigma$ is
\[
EV_p^{\sigma}(I_p)=\sum_{h\in I_p}\pi^{\sigma}(h\mid I_p)\sum_{z\in\mathcal Z,h\sqsubseteq z}\pi^{\sigma}(z\mid h)u_p(z).
\]
The best response to $\sigma_{-p}$ is $\text{BR}(\sigma_{-p}) = \arg\max_{\sigma_p'} u_p(\sigma_p', \sigma_{-p})$.
A Nash equilibrium $\sigma^*$ satisfies $\forall p,\; u_p(\sigma^*) = \max_{\sigma_p} u_p(\sigma_p, \sigma_{-p}^*)$.
The exploitability of $\sigma$ is $\text{Expl}(\sigma) = \frac{1}{2} \sum_{p} u_p(\text{BR}(\sigma_{-p}), \sigma_{-p})$.

Counterfactual Regret Minimization (CFR) \citep{zinkevich2007regret} iteratively computes regrets $r^t(I,a)=v_p^{\sigma^t}(I,a)-v_p^{\sigma^t}(I)$, regret sums $R^T(I, a) = \sum_{t=1}^{T} r^t(I, a)$ and derives strategies via regret matching: $\sigma^{t+1}(I,a) = R_+^t(I,a) / \sum_{a'} R_+^t(I,a')$, where $R_+ = \max(0, R)$.
The average strategy $\bar{\sigma}^T$ converges to an $\varepsilon$-Nash equilibrium.

A bucket mapping $\phi: \mathcal{I}_p^0 \to \{1, \ldots, K\}$ assigns each root information set to one of $K$ buckets.
All downstream information sets derived from the same root information set are mapped to the same bucket and share a single strategy, reducing the effective strategy space from $|\mathcal{I}_p^0|$ to $K$.
In the abstracted game, CFR operates over buckets rather than individual information sets.

\section{Method}
\label{sec:method}

Given a game $G$, bucket count $K$, and warm-up iterations $W$, WEVA proceeds in four steps:
\begin{enumerate}
    \item \textbf{Warm-up:} Run $W$ iterations of CFR on the full (unabstracted) game $G$, producing an average strategy profile $\overline{\sigma}^W$.
    \item \textbf{EV extraction:} For each root information set $I_p^0 \in \mathcal{I}_p^0$ and each decision node $n$ reachable from the root, compute the expected value $EV_p^{\overline{\sigma}^W}(I_p^0{\downarrow}n)$ of the corresponding downstream information set.
    \item \textbf{Clustering:} Construct a feature vector $\mathbf{f}(I_p^0)$ from the extracted EVs, then apply $k$-means++ to partition root information sets into $K$ buckets.
    \item \textbf{Solving:} Use the bucket mapping to define an abstract game $G'$ and solve $G'$ with CFR for $T$ iterations.
\end{enumerate}
The warm-up phase uses no abstraction (each root information set is its own bucket), so the intermediate strategy captures the full game dynamics. The total computational cost is $W$ iterations on the full game plus $T$ iterations on the abstract game. Algorithm~\ref{alg:warm-upev} summarizes the procedure.

\begin{algorithm}[t]
\caption{WEVA Abstraction}
\label{alg:warm-upev}
\KwIn{Game $G$, bucket count $K$, warm-up iterations $W$, solve iterations $T$, depth weights $\{w_d\}$}
\KwOut{Bucket mapping $\phi$, strategy profile $\bar{\sigma}^T$}
$\overline{\sigma}^W \leftarrow$ Run $W$ iterations of CFR on full game $G$\;
\For{$p\in\mathcal N$}{
\For{each $I_p^0 \in \mathcal{I}_p^0$}{
  \For{each decision node $n\in G$}{
    Compute $EV_p^{\overline{\sigma}^W}(I_p^0{\downarrow}n)$\;
  }
  $\mathbf{f}(I_p^0) \leftarrow [w_{d_0} \cdot EV_p^{\overline{\sigma}^W}(I_p^0{\downarrow}n_0),\; \ldots,\; w_{d_{N-1}} \cdot EV_p^{\overline{\sigma}^W}(I_p^0{\downarrow}n_{N-1})]$\;
}
$\phi_p \leftarrow k\text{-means++}(\{\mathbf{f}(I_p^0)\}_{I_p^0 \in \mathcal{I}_p^0},\, K)$\;
}
$\phi\leftarrow \{\phi_p\}_{p\in\mathcal N}$\;
Construct abstract game $G'$ using bucket mapping $\phi$\;
$\bar{\sigma}^T \leftarrow$ Run $T$ iterations of CFR on $G'$\;
\Return{$\phi, \bar{\sigma}^T$}
\end{algorithm}

\paragraph{Root-EV (ev-root).}
The simplest variant uses only the root-node EV as a one-dimensional feature:
\begin{equation}
    f_{\text{root}}(I_p^0) = EV_p^{\overline{\sigma}^W}(I_p^0{\downarrow}n_{\text{root}})
\end{equation}
where $n_{\text{root}}$ is the root decision node (so $I_p^0{\downarrow}n_{\text{root}} = I_p^0$). Clustering is performed via one-dimensional $k$-means. This feature is semantically similar to equity but derived from actual game play rather than showdown evaluation, making it applicable to games without standard showdown structures.

\paragraph{Multi-Node Depth-Weighted EV (ev-nd).}
Two root information sets with identical root-node EVs may behave very differently at deeper decision nodes. To capture these strategic differences, ev-nd constructs a multi-dimensional feature vector incorporating downstream EVs from \emph{all} reachable decision nodes, weighted by depth. The idea of using multi-node EV features for abstraction was first introduced by \citet{li2025evpa}, who train neural networks on billions of samples to estimate per-node EVs, incurring substantial offline cost. WEVA instead extracts these features directly from a short CFR warm-up, avoiding the need for any offline training while providing more accurate EV estimates from actual game-play dynamics.

Let $\{n_0, n_1, \ldots, n_{N-1}\}$ be the decision nodes at depths $\{d_0, d_1, \ldots, d_{N-1}\}$. The feature vector for root information set $I_p^0$ is:
\begin{equation}
    \mathbf{f}_{\text{nd}}(I_p^0) = \big[w_{d_0} \cdot EV_p^{\overline{\sigma}^W}(I_p^0{\downarrow}n_0),\; w_{d_1} \cdot EV_p^{\overline{\sigma}^W}(I_p^0{\downarrow}n_1),\; \ldots,\; w_{d_{N-1}} \cdot EV_p^{\overline{\sigma}^W}(I_p^0{\downarrow}n_{N-1})\big]
\end{equation}
where $w_d$ is a manually specified depth-dependent weight (see Appendix~\ref{app:details} for details). The high weight at depth 0 reflects that root-level value is the most informative signal, while deeper nodes provide diminishing but still useful refinement. Clustering is performed via $k$-means++ \citep{DBLP:conf/soda/ArthurV07}.

\paragraph{Computational cost.}
The warm-up phase runs $W$ iterations of CFR on the full game with $|\mathcal{I}_p^0|$ root information sets and $D$ decision nodes, costing $O(W \cdot |\mathcal{I}_p^0| \cdot D)$ per iteration. The subsequent abstract solve runs $T$ iterations with $K$ buckets, costing $O(T \cdot K \cdot D)$ per iteration. The relative overhead of warm-up is:
\begin{equation}
    \text{overhead} = \frac{W \cdot |\mathcal{I}_p^0|}{T \cdot K}
\end{equation}
For typical settings ($W{=}10$, $T{=}2{,}000$, $K{=}200$, $|\mathcal{I}_p^0|{=}1{,}081$ in river HUNL), this ratio is approximately $2.7\%$.

\section{Experiments}
\label{sec:experiments}

\subsection{Setup}

We evaluate WEVA on three structurally diverse benchmarks: \textbf{HUNL Endgame}, \textbf{Double-Board HUNL}, and a \textbf{Random Game}, whose detailed definitions appear in Appendix~\ref{app:games}. These games span a wide spectrum: HUNL has strong domain features, Double-Board breaks single-dimensional hand strength, and the Random Game eliminates all domain structure.

\begin{figure*}[!ht]
\centering
\includegraphics[width=\textwidth]{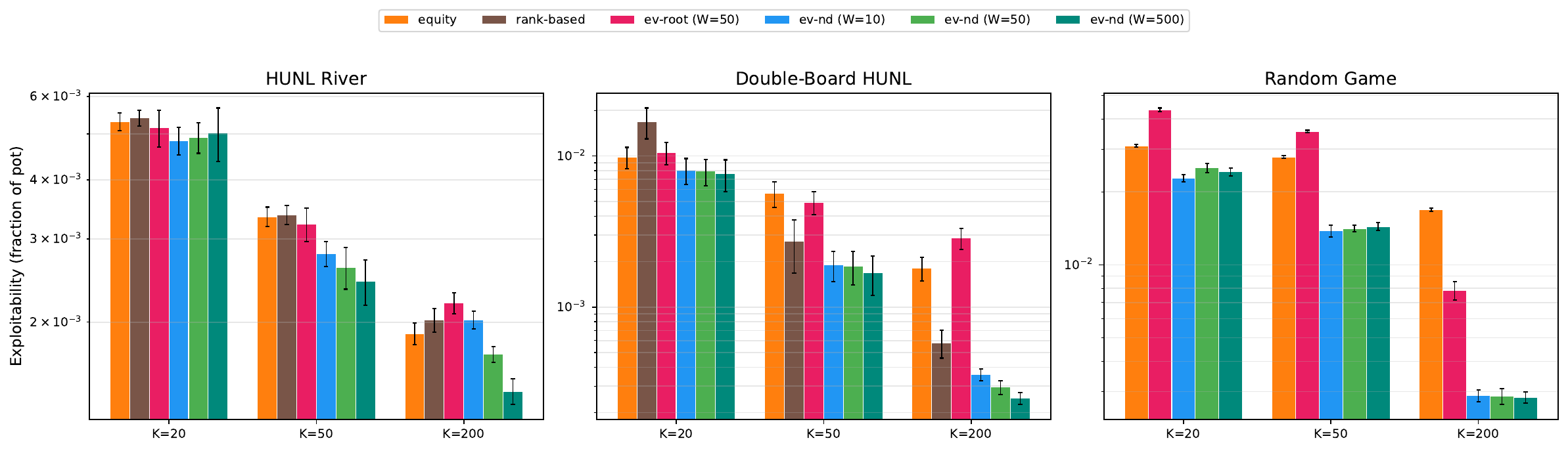}
\caption{Final exploitability (log scale, lower is better) at $T{=}2{,}000$ for all methods on the three games with PCFR+. The rank-based method means rank-based abstraction in HUNL Endgame and rank-2d abstraction in Double-Board HUNL.
}
\label{fig:main_bar}
\end{figure*}

Our implementation builds on OpenSpiel \citep{lanctot2019openspiel} and LiteEFG \citep{liu2024liteefg}. For each benchmark, we generate $10$ random boards and average results. We use PCFR+ as the primary CFR variant and DCFR results are reported in Appendix~\ref{app:dcfr}. The CFR iteration budget $T$ is $2{,}000$.  Bucket counts are $K \in \{20, 50, 200\}$ and warm-up iterations are $W \in \{10, 50, 500\}$. Baseline methods include \textbf{equity} (showdown win probability) and \textbf{rank} (hand strength ranking). For Double-Board HUNL, we additionally test \textbf{rank-2d}, a game-specific baseline that clusters hands using a 2D feature vector of per-board ranks via $k$-means.

Figure~\ref{fig:main_bar} summarizes final exploitability across all methods, games, and $K$ values for PCFR+. Tables~\ref{tab:hunl}--\ref{tab:random_game} provide the corresponding numerical values, which we analyze game by game below.

\subsection{HUNL Endgame Results}

Table~\ref{tab:hunl} shows results on HUNL endgames. Equity is a strong baseline in this setting, as relative hand strength directly determines showdown outcomes. Nevertheless, ev-nd consistently outperforms equity across all $K$ values:

\begin{table}[h]
\centering
\caption{Exploitability (fraction of pot) on HUNL Endgame with PCFR+. Lower is better.}
\label{tab:hunl}
\begin{tabular}{@{}l c c c@{}}
\toprule
Method & $K{=}20$ & $K{=}50$ & $K{=}200$ \\
\midrule
equity & $.00531 \pm .00023$ & $.00334 \pm .00016$ & $.00189 \pm .00010$ \\
rank & $.00541 \pm .00021$ & $.00337 \pm .00015$ & $.00202 \pm .00012$ \\
\midrule
ev-root ($W{=}50$) & $.00515 \pm .00046$ & $.00323 \pm .00026$ & $.00220 \pm .00011$ \\
ev-nd ($W{=}10$) & $\mathbf{.00484} \pm .00032$ & $.00279 \pm .00017$ & $.00202 \pm .00009$ \\
ev-nd ($W{=}50$) & $.00492 \pm .00036$ & $.00261 \pm .00027$ & $.00171 \pm .00007$ \\
ev-nd ($W{=}500$) & $.00502 \pm .00065$ & $\mathbf{.00244} \pm .00027$ & $\mathbf{.00143} \pm .00009$ \\
\bottomrule
\end{tabular}
\end{table}

At $K{=}20$, ev-nd reduces exploitability versus equity by $9\%$, $7\%$, and $5\%$ at $W{\in}\{10,50,500\}$, and versus rank by $11\%$, $9\%$, and $7\%$. At $K{=}50$, the gains grow with $W$: $16\%$, $22\%$, and $27\%$ versus equity and rank. At $K{=}200$, the effect of $W$ is the most pronounced: $W{=}10$ yields a slight regression ($-7\%$ vs equity, $0\%$ vs rank), while $W{=}50$ and $W{=}500$ recover strongly, reaching $10\%$ and $24\%$ versus equity (and $15\%$, $29\%$ versus rank). On HUNL Endgame, a richer warm-up signal is therefore necessary to distinguish 200 buckets, while at coarser granularities the $W{=}10$ signal already suffices. The ev-root variant performs at the same level against equity or rank, confirming that root-level information alone is insufficient.

The relatively modest gap between equity/rank-based and ev-nd on HUNL Endgame is expected: in HUNL, showdown payoffs are fully determined by a monotonic hand-strength ordering, so equity and rank-based already capture the dominant structure. The marginal advantage of ev-nd comes from capturing \emph{blocker effects} that one-dimensional features miss. For example, on a board with three suited cards (e.g., $7\heartsuit\,9\heartsuit\,Q\heartsuit\,3\diamondsuit\,K\clubsuit$), a player holding $A\heartsuit\,2\spadesuit$ has low equity (ace-high) but uniquely blocks the opponent's nut flush ($A\heartsuit$-$x\heartsuit$). Knowing this, the player can bluff more aggressively at certain decision nodes because the opponent can never hold the strongest flush. Ev-nd's multi-node features detect such differences in downstream EV across nodes, assigning blocker hands to different buckets from hands with similar equity but no blocking effect.

\subsection{Double-Board HUNL Results}

Double-Board HUNL breaks the single-dimensional hand-strength assumption. As shown in Table~\ref{tab:double_board}, 1D rank abstraction collapses entirely, as a single ordering cannot capture two-dimensional hand strength. The game-specific rank-2d baseline, which clusters on per-board rank vectors, performs much better but still falls short of ev-nd:

\begin{table}[h]
\centering
\caption{Exploitability on Double-Board HUNL with PCFR+. $^\dagger$Game-specific method using per-board rank features.}
\label{tab:double_board}
\begin{tabular}{@{}l c c c@{}}
\toprule
Method & $K{=}20$ & $K{=}50$ & $K{=}200$ \\
\midrule
equity & $.00983 \pm .00157$ & $.00564 \pm .00107$ & $.00181 \pm .00032$ \\
rank & $.07261 \pm .00821$ & $.05561 \pm .00715$ & $.01698 \pm .00277$ \\
rank-2d$^\dagger$ & $.01688 \pm .00390$ & $.00272 \pm .00104$ & $.00058 \pm .00012$ \\
\midrule
ev-root ($W{=}50$) & $.01053 \pm .00180$ & $.00494 \pm .00086$ & $.00287 \pm .00046$ \\
ev-nd ($W{=}10$) & $.00804 \pm .00157$ & $.00190 \pm .00043$ & $.00036 \pm .00003$ \\
ev-nd ($W{=}50$) & $.00791 \pm .00158$ & $.00186 \pm .00046$ & $.00029 \pm .00003$ \\
ev-nd ($W{=}500$) & $\mathbf{.00762} \pm .00181$ & $\mathbf{.00169} \pm .00049$ & $\mathbf{.00025} \pm .00002$ \\
\bottomrule
\end{tabular}
\end{table}

At $K{=}20$, ev-nd reduces exploitability versus equity by $18\%$, $20\%$, and $22\%$ at $W{\in}\{10,50,500\}$, and versus rank-2d by $52\%$, $53\%$, and $55\%$. At $K{=}50$, the gains versus equity grow sharply: $66\%$, $67\%$, and $70\%$. And the improvements against game-specific rank-2d are $30\%$, $32\%$ and $38\%$. At $K{=}200$, the gains against equity are $80\%$, $84\%$ and $86\%$, and the improvements against rank-2d are $38\%$, $50\%$ and $57\%$. The  rank-2d abstraction, a hand-crafted multi-board feature, is better than equity at $K{=}50$ and $K{=}200$, however, still weaker than ev-nd. Hands strong on one board and weak on the other behave differently at various decision nodes, and ev-nd's multi-node features distinguish them already within $10$ CFR iterations.

\subsection{Random Game Results}

The Random Game provides the most challenging test: payoffs are random and independent across nodes, so no domain feature has structural meaning. As shown in Table~\ref{tab:random_game}, equity-based abstraction struggles, while ev-nd works really well.

\begin{table}[h]
\centering
\caption{Exploitability on Random Game with PCFR+.}
\label{tab:random_game}
\begin{tabular}{@{}l c c c@{}}
\toprule
Method & $K{=}20$ & $K{=}50$ & $K{=}200$ \\
\midrule
equity & $.03100 \pm .00044$ & $.02785 \pm .00031$ & $.01685 \pm .00025$ \\
\midrule
ev-root ($W{=}50$) & $.04363 \pm .00076$ & $.03554 \pm .00036$ & $.00784 \pm .00068$ \\
ev-nd ($W{=}10$) & $\mathbf{.02284} \pm .00080$ & $\mathbf{.01381} \pm .00077$ & $.00288 \pm .00016$ \\
ev-nd ($W{=}50$) & $.02513 \pm .00108$ & $.01411 \pm .00043$ & $.00287 \pm .00022$ \\
ev-nd ($W{=}500$) & $.02417 \pm .00089$ & $.01440 \pm .00054$ & $\mathbf{.00284} \pm .00015$ \\
\bottomrule
\end{tabular}
\end{table}

At $K{=}20$, ev-nd reduces exploitability versus equity by $26\%$, $19\%$, and $22\%$ at $W{\in}\{10,50,500\}$; notably, $W{=}10$ is the strongest setting. At $K{=}50$: $51\%$, $49\%$, and $48\%$. At $K{=}200$, the improvement saturates at $83\%$ regardless of $W$. Unlike HUNL and Double-Board HUNL Endgames, larger warm-up brings essentially no additional benefit on the Random Game, suggesting that its EV structure stabilizes extremely quickly. 

\subsection{Key Findings}

Figure~\ref{fig:warmup_ablation} shows the normalized exploitability against equity baseline across all methods and bucket settings. WEVA consistently outperforms all baselines in all games and all bucket settings. Here are our key findings:

\begin{figure}[h]
\centering
\includegraphics[width=\columnwidth]{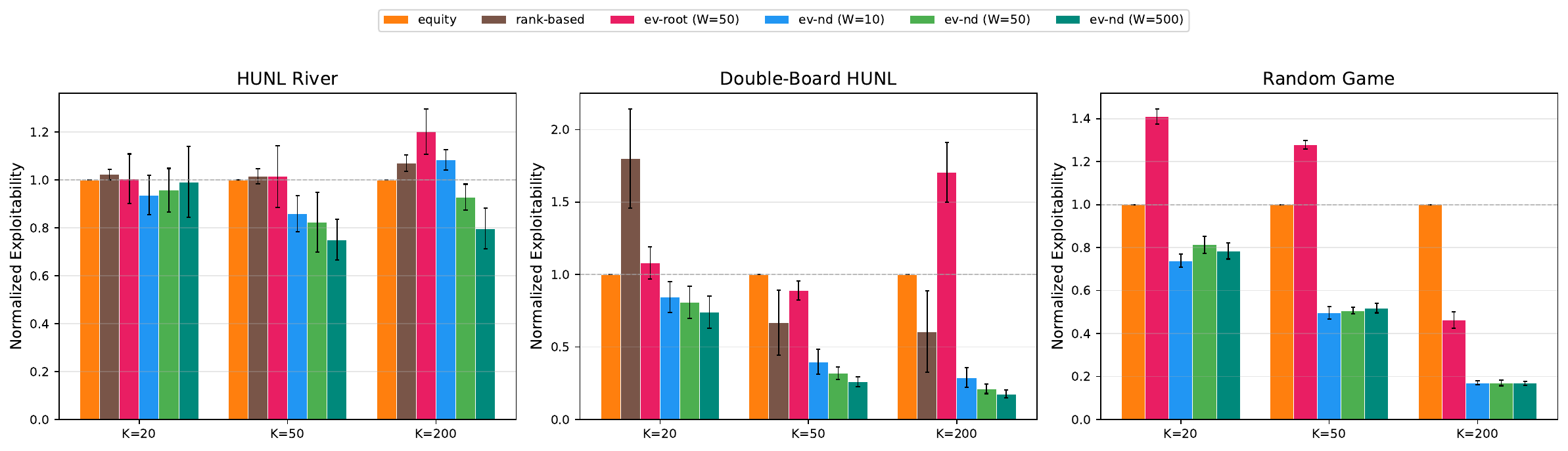}
\caption{Normalized exploitability (ratio to equity baseline) across all methods and $K \in \{20, 50, 200\}$. For each board, we compute exploitability(method)/exploitability(equity). Values below $1.0$ indicate improvement over equity.}
\label{fig:warmup_ablation}
\end{figure}

\begin{itemize}
    \item \textbf{$W{=}10$ is already effective at small $K$}: Across all three games at $K{\in}\{20,50\}$, the improvement over equity with $W{=}10$ is within a few percentage points of $W{=}500$, and on the Random Game $W{=}10$ even dominates. Warm-up is essentially free in this regime.
    \item \textbf{Larger $W$ may help at fine granularities}: At $K{=}200$, $W{=}500$ surpasses $W{=}10$ on HUNL Endgame ($-7\%{\to}24\%$ vs equity) and Double-Board HUNL ($80\%{\to}86\%$ vs equity, $38\%{\to}57\%$ vs rank-2d). The Random Game saturates at $83\%$ regardless of $W$, indicating that its EV structure stabilizes within the first $10$ iterations.

\end{itemize}
This yields a practical rule: at small $K$, $W{=}10$ already captures the relevant ordinal structure and additional warm-up iterations provide diminishing returns; at large $K$, the abstract-game solve dominates total runtime, and the extra cost of a longer warm-up phase can be weighed against the resulting accuracy improvements.

\section{Conclusion}
\label{sec:conclusion}

We presented WEVA, a lightweight information abstraction method that extracts EV features from a short CFR warm-up phase. The key variant, ev-nd, constructs multi-dimensional depth-weighted features from all decision nodes, capturing strategic differences that single-dimensional features miss. WEVA requires no domain knowledge, no neural network training, and incurs very small additional computational cost. The previous abstract work was often closed source and domain specific. We have provided the first effective, efficient and general open-source implementation of information abstraction in IIEFGs, which has significant implications for future research in the field of abstraction. Experiments on three diverse games show that WEVA consistently and often dramatically outperforms traditional abstraction methods, reducing exploitability by up to over $80\%$.  Surprisingly, we find that as few as $W{=}10$ warm-up iterations suffice, which makes the approach practical for large-scale game solving.


\bibliographystyle{plainnat}
\bibliography{references}

\appendix

\newpage
\section{DCFR Results}
\label{app:dcfr}

\begin{figure*}[!ht]
\centering
\includegraphics[width=\textwidth]{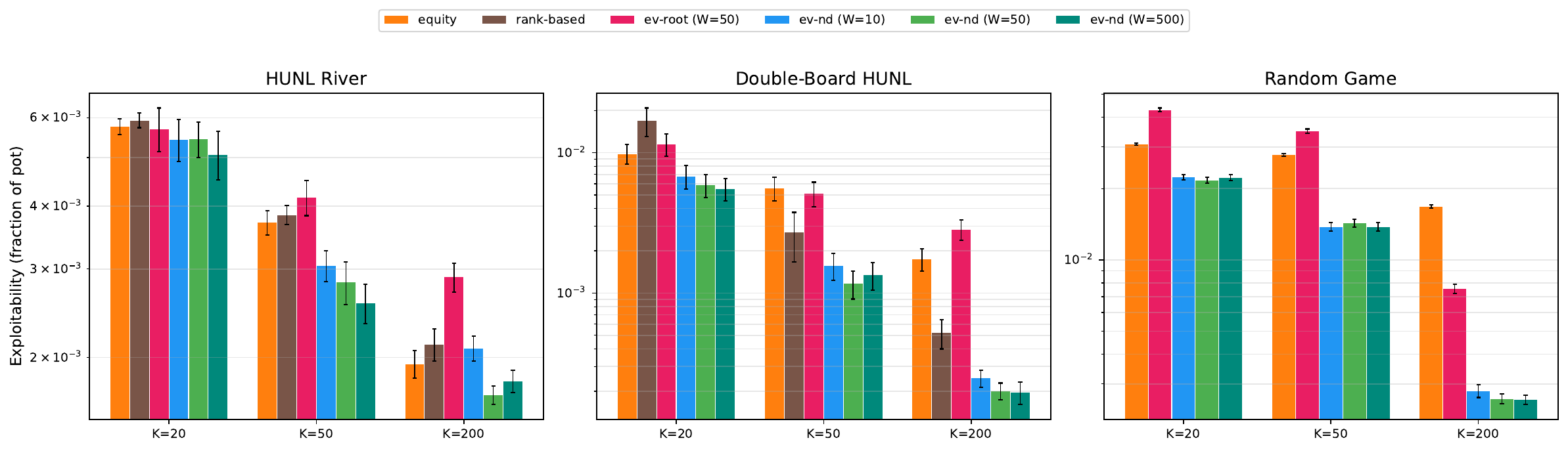}
\caption{Final exploitability at $T{=}2{,}000$ for all methods on the three games with DCFR.}
\label{fig:dcfr_bar}
\end{figure*}

We repeat all experiments using DCFR  as the solver. Figure~\ref{fig:dcfr_bar} summarizes final exploitability across all methods, games, and $K$ values for DCFR. Tables~\ref{tab:hunl_dcfr}--\ref{tab:random_game_dcfr} provide the corresponding numerical values. Results are consistent with PCFR+: ev-nd outperforms all baselines, and the relative improvements are similar in magnitude.

\begin{table}[!ht]
\centering
\caption{Exploitability on HUNL Endgame with DCFR. }
\label{tab:hunl_dcfr}
\begin{tabular}{@{}l c c c@{}}
\toprule
Method & $K{=}20$ & $K{=}50$ & $K{=}200$ \\
\midrule
equity & $.00577 \pm .00021$ & $.00371 \pm .00021$ & $.00194 \pm .00012$ \\
rank & $.00593 \pm .00020$ & $.00384 \pm .00017$ & $.00212 \pm .00016$ \\
\midrule
ev-root ($W{=}50$) & $.00570 \pm .00057$ & $.00417 \pm .00034$ & $.00289 \pm .00019$ \\
ev-nd ($W{=}10$) & $.00543 \pm .00052$ & $.00305 \pm .00021$ & $.00209 \pm .00012$ \\
ev-nd ($W{=}50$) & $.00544 \pm .00044$ & $.00282 \pm .00027$ & $.00168 \pm .00007$ \\
ev-nd ($W{=}500$) & $\mathbf{.00507} \pm .00056$ & $\mathbf{.00256} \pm .00023$ & $\mathbf{.00180} \pm .00009$ \\
\bottomrule
\end{tabular}
\end{table}

\begin{table}[!ht]
\centering
\caption{Exploitability on Double-Board HUNL with DCFR. }
\label{tab:double_board_dcfr}
\begin{tabular}{@{}l c c c@{}}
\toprule
Method & $K{=}20$ & $K{=}50$ & $K{=}200$ \\
\midrule
equity & $.00984 \pm .00158$ & $.00560 \pm .00107$ & $.00175 \pm .00032$ \\
rank & $.07190 \pm .00825$ & $.05547 \pm .00712$ & $.01691 \pm .00277$ \\
rank-2d$^\dagger$ & $.01690 \pm .00391$ & $.00271 \pm .00105$ & $.00052 \pm .00012$ \\
\midrule
ev-root ($W{=}50$) & $.01152 \pm .00208$ & $.00513 \pm .00103$ & $.00285 \pm .00047$ \\
ev-nd ($W{=}10$) & $.00679 \pm .00128$ & $.00158 \pm .00034$ & $.00025 \pm .00003$ \\
ev-nd ($W{=}50$) & $.00589 \pm .00109$ & $.00117 \pm .00026$ & $\mathbf{.00020} \pm .00003$ \\
ev-nd ($W{=}500$) & $\mathbf{.00552} \pm .00100$ & $\mathbf{.00135} \pm .00030$ & $\mathbf{.00020} \pm .00004$ \\
\bottomrule
\end{tabular}
\end{table}

\begin{table}[!ht]
\centering
\caption{Exploitability on Random Game with DCFR. }
\label{tab:random_game_dcfr}
\begin{tabular}{@{}l c c c@{}}
\toprule
Method & $K{=}20$ & $K{=}50$ & $K{=}200$ \\
\midrule
equity & $.03083 \pm .00036$ & $.02776 \pm .00032$ & $.01682 \pm .00024$ \\
\midrule
ev-root ($W{=}50$) & $.04299 \pm .00080$ & $.03496 \pm .00074$ & $.00757 \pm .00035$ \\
ev-nd ($W{=}10$) & $\mathbf{.02238} \pm .00056$ & $\mathbf{.01380} \pm .00059$ & $.00280 \pm .00017$ \\
ev-nd ($W{=}50$) & $.02173 \pm .00064$ & $.01433 \pm .00055$ & $.00260 \pm .00013$ \\
ev-nd ($W{=}500$) & $.02232 \pm .00062$ & $.01380 \pm .00059$ & $\mathbf{.00257} \pm .00012$ \\
\bottomrule
\end{tabular}
\end{table}

\section{Game Definitions}
\label{app:games}

\paragraph{HUNL Endgame.} A standard poker river subgame. Five community cards are dealt uniformly at random. Each player's hand space consists of all $\binom{47}{2} = 1{,}081$ possible 2-card combinations from the remaining deck, with card blocking enforced (no two players share the same card). At showdown, the player with the stronger 5-card poker hand wins the pot.

\paragraph{Double-Board HUNL.} Ten community cards are dealt and split into two independent boards of five cards each. Each player holds 2 hole cards from the remaining 42 cards ($\binom{42}{2} = 861$ possible hands). At showdown, each board is evaluated independently using standard poker hand rankings, and the pot is split equally between the winners of each board.

\paragraph{Random Game.} $500$ abstract hands per player with no card structure. Terminal payoffs are drawn i.i.d.\ from $\{+1, -1\}$ independently for each terminal node and each hand matchup, forming an antisymmetric payoff matrix.

\section{Discussion}
\label{sec:discussion}

\paragraph{Scope and limitations.}
WEVA is designed for IIEFGs where private information is fixed at the start of the game, fully enumerable, and does not change during play. The game tree contains no private actions and no intermediate chance nodes that alter private information. These assumptions hold for river subgames of poker (where hole cards are fixed and all community cards are already dealt) and many other endgame settings, but they exclude games with ongoing private information revelation (e.g. Mahjong) or games where the private type space is too large to enumerate exhaustively (e.g. dark chess). Additionally, the warm-up phase runs CFR on the full (unabstracted) game, which requires the full information-set space to fit in memory. For games where the unabstracted game is too large to solve directly (e.g., HUNL), WEVA cannot be applied to the entire game tree directly.

\paragraph{Future works.} A natural extension is to combine WEVA with belief-based depth-limited subgame solving \citep{moravcik2017deepstack,brown2020combining,li2026turborebel}. In this paradigm, the solver decomposes the game into subgames at runtime and solves each subgame independently. WEVA can serve as the abstraction layer within each subgame: upon reaching a new subgame, a short warm-up produces EV-based features for that subgame's information sets, which are then clustered into buckets for efficient solving. This \emph{online abstraction} approach has a key advantage over prior imperfect-recall abstraction methods \citep{ganzfried2014potential} and neural-network based abstraction methods \citep{brown2019deep,li2025evpa}, which require extensive offline precomputation. With WEVA, abstraction is computed on-the-fly with negligible overhead, enabling real-time deployment in large-scale games such as HUNL.

Several directions merit exploration as well. First, applying WEVA to games with larger information-set spaces, such as Pot Limit Omaha and multi-player No-Limit Hold'em \citep{brown2019superhuman}, would test scalability. Second, alternative warm-up features beyond EV, such as reach probabilities or regret distributions, may provide complementary clustering signals. Third, investigating the interaction between warm-up length $W$ and game complexity could yield adaptive schedules that allocate warm-up budget where it matters most.

\section{Experimental Details}
\label{app:details}

Table~\ref{tab:hyperparams} summarizes the hyperparameters used in all experiments.

\begin{table}[h]
\centering
\caption{Hyperparameters for all experiments.}
\label{tab:hyperparams}

\begin{tabular}{@{}l l@{}}
\toprule
Parameter & Value \\
\midrule
Number of random boards & 10 \\
Seed & 42+board\_id\\
Bet sizes & $0.5\times$ pot, $1\times$ pot, $2\times$ pot \\
DCFR ($\alpha, \beta, \gamma$) & $(1.5,\; 0,\; 2)$ \\
Solve iterations $T$ & $2{,}000$ \\
Warm-up iterations $W$ & $\{10, 50, 500\}$ \\
Bucket counts $K$ & $\{20, 50, 200\}$ \\
$k$-means max iterations & $8$ \\
Depth attention weights $w_0, w_1, w_2, \ldots$ & $5.0,\; 1.0,\; 0.5,\; 0.25,\; 0.15,\; 0.1,\; 0.07,\; 0.05,\; \ldots$ \\
\bottomrule
\end{tabular}
\end{table}

\subsection{Per-Board Variance}
\label{app:variance}

\begin{figure}[h]
\centering
\includegraphics[width=\columnwidth]{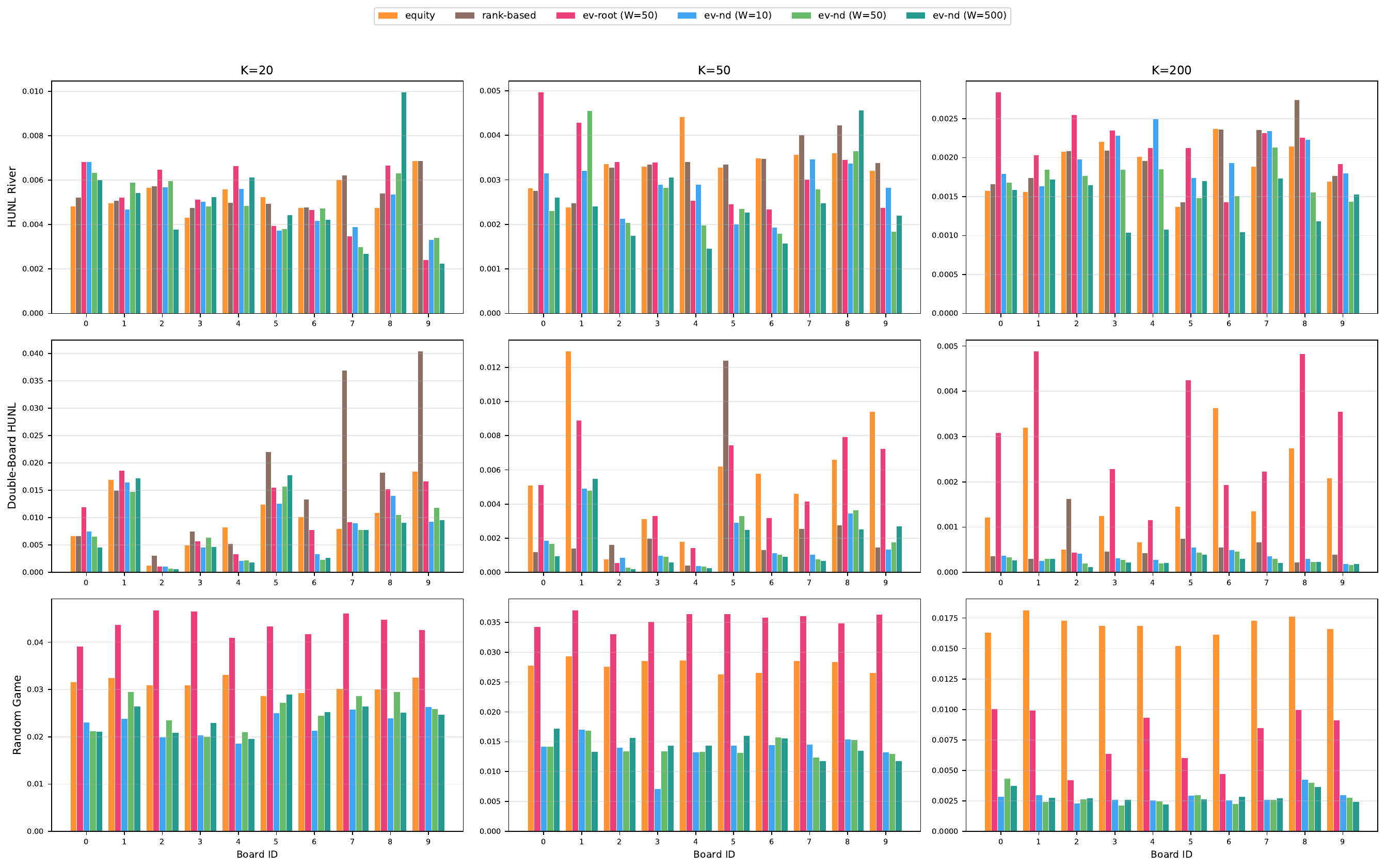}
\caption{Per-board exploitability for all abstraction methods across 10 boards, three games, and three $K$ values. Each group of bars represents one board, and different colors correspond to different methods. 
}
\label{fig:per_board}
\end{figure}

Figure~\ref{fig:per_board} shows per-board exploitability for all abstraction methods across 10 boards. Each board exhibits different characteristics: on some boards one method performs best, while on others a different method dominates. This variation reflects the inherent diversity of random board draws. Importantly, ev-nd variants achieve consistently low exploitability across the majority of boards, confirming that the improvements in the main tables are not driven by outliers.

\subsection{Convergence Curves}
\label{app:convergence}

\begin{figure*}[h]
\centering
\includegraphics[width=\textwidth]{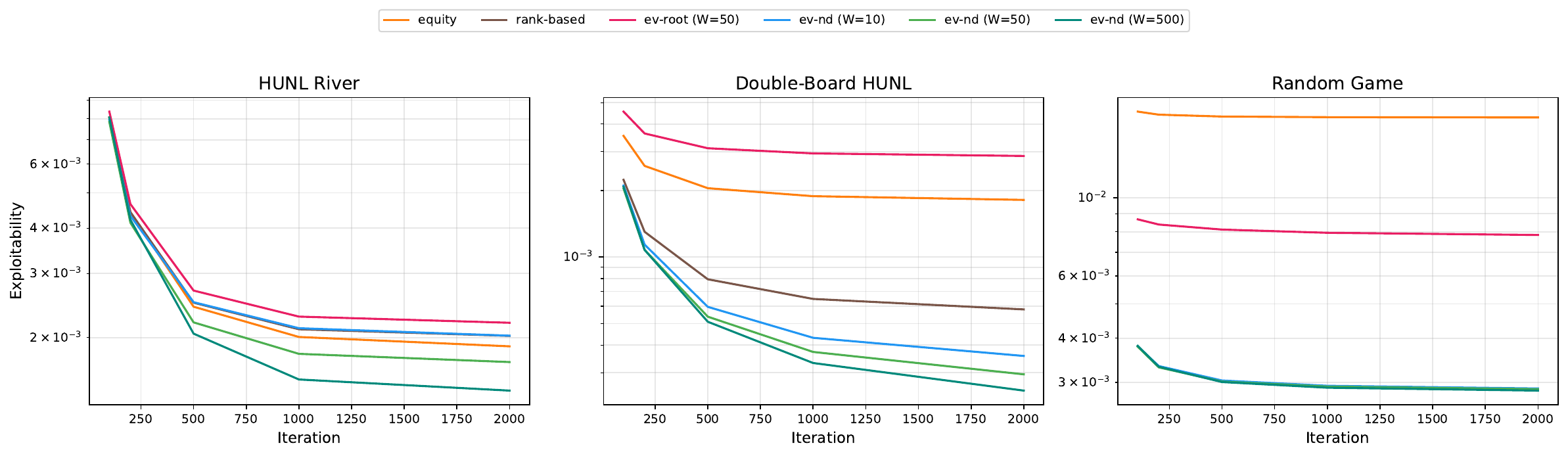}
\caption{Convergence curves at $K{=}200$ across three games using PCFR+ with $T=2{,}000$ iterations.
}
\label{fig:convergence_combined}
\end{figure*}

Figure~\ref{fig:convergence_combined} shows convergence curves at bucket count $K{=}200$ across all three games using PCFR+ with $T=2{,}000$ iterations. The gains in Tables~\ref{tab:hunl}--\ref{tab:random_game} are not only in the final exploitability: ev-nd converges to lower exploitability across all games and the advantage emerges within the first few hundred iterations, widening thereafter.

\end{document}